\documentclass{aa}
\usepackage{graphicx}

\def\deg{$^\circ$}

\def\arcsec{$^{\prime\prime\,}$}
\def\arcmin{$^{\prime\,}$}

\def\a{$^{\mbox{\small a}}$}
\def\b{$^{\mbox{\small b}}$}
\def\c{$^{\mbox{\small c}}$}
\def\d{$^{\mbox{\small d}}$}
\def\e{$^{\mbox{\small e}}$}
\def\f{$^{\mbox{\small f}}$}
\def\*{$^{*}$}

\begin{document}

\sloppypar
 
\title{INTEGRAL observations of five sources in the Galactic Center region}   
 
\author{A.Lutovinov \inst{1}, M.Revnivtsev\inst{1,2}, S.Molkov \inst{1},
  R.Sunyaev\inst{1,2}}     

\offprints{lutovinov@hea.iki.rssi.ru} 

\institute{Space Research Institute, Russian Academy of Sciences,
              Profsoyuznaya 84/32, 117810 Moscow, Russia
\and
Max-Planck-Institute f\"ur Astrophysik, Karl-Schwarzschild-Str. 1, D-85740
Garching bei M\"unchen, Germany
}

\date{Received /Accepted}

        \authorrunning{Lutovinov et al.}
        \titlerunning{INTEGRAL observations of five sources} 
        
        \abstract{A number of new X-ray sources (IGR J17091-3624, IGR/XTE
          J17391-3021, IGR J17464-3213 (= XTE J17464-3213 = H 1743-322), IGR
          J17597-2201, SAX/IGR J18027-2017) have been observed with the
          INTEGRAL observatory during ultra deep exposure of the Galactic
          Center region in August-September 2003.  Most of them were
          permanently visible by the INTEGRAL at energies higher than $\sim
          20$ keV, but IGR/XTE J17391-3021 was observed only during its
          flaring activity with a flux maximum of $\sim120$ mCrab. IGR
          J17091-3624, IGR J17464-3213 and IGR J17597-2201 were detected up
          to $\sim 100$-$150$ keV. In this paper we present the analysis of
          INTEGRAL observations of these sources to determine the nature of
          these objects. We conclude that all of them have a galactic
          origin. Two sources are black hole candidates (IGR J17091-3624 and
          IGR J17464-3213), one is an LMXB neutron star binary (presumably
          an X-ray burster) and two other sources (IGR J17597-2201 and
          SAX/IGR J18027-2017) are neutron stars in high mass binaries; one
          of them (SAX/IGR J18027-2017) is an accreting X-ray pulsar.
           }

   \maketitle
%

\section*{Introduction}

About 12 new hard X-ray sources have been discovered with the INTEGRAL
observatory. Most of them are concentrated towards the Galactic plane,
Galactic Center and spiral arm tangents (see \cite{rev04a}, \cite{mol04},
\cite{tom04}, \cite{bird04} and references therein). Such a concentration
strongly suggests that the majority of these sources have a galactic origin.

An ultra deep survey of the Galactic Center region, performed by the
INTEGRAL in 2003 provided very important information about the population of
Galactic X-ray sources in the bulge and revealed a number of transients in
this field. A previous deep survey of this region in hard X-ray/soft
gamma-rays was made by the SIGMA telescope aboard the GRANAT orbital
observatory with a sensitivity of 3-5 mCrab in the 40-150 keV energy band
and allowed us to detect 20 sources including transients (\cite{chur94},
\cite{rev04b}). Recent analysis of the images of the Galactic Center region
obtained by INTEGRAL/IBIS in the energy range 18-60 keV with a typical
sensitivity of 1-2 mCrab revealed 60 sources of different natures: low-mass
X-ray binaries, X-ray pulsars, black-hole candidates, cataclysmic variables,
extragalactic sources etc. (\cite{rev04a}). The nature of a significant part
of the detected sources is still unclear. Out of 60 detected sources six new
sources were detected during INTEGRAL observations of the Galactic Center
region. Five of them were statistically significantly detected on the
averaged map of the region, while IGR J17544-2619 demonstrated a very
peculiar behavior (\cite{sun03a}, \cite{gre03}, \cite{intz04},
\cite{gon04}). It was detected only during a few ksec of observations and we
do not discuss it here.

In this paper we summarize the information about properties of new INTEGRAL
sources and try to determine the nature of these sources. 

\section*{Observations and data reduction}

\begin{table*}[t]
\caption{Journal of observations \label{tab1}}   
\centering

\small{
\tabcolsep=1.3mm
\begin{tabular}{l|l|l|c|c|c|c|l}
\hline\hline
Source &R.A.\a&Dec.\a&\multicolumn{2}{c|}{2003} & \multicolumn{2}{c|}{2004} &Source Type\\  
\cline{4-7}&(2000)&(2000)& Exp., ksec\b & Flux, mCrab\c& Exp., ksec & Flux, mCrab& \\
\hline
IGR J17091-3624 &17$^h$09$^m$06$^s$&36$^\circ$24\arcmin38\arcsec& 920 & 7.6$\pm$0.2 & 96 & 16.8$\pm$0.6             &BHC\\
IGR/XTE J17391-3021 &17$^h$39$^m$06$^s$&30$^\circ$21\arcmin30\arcsec& 2055 & 1.6$\pm$0.1 (123$\pm$3)\d & 236 & 1.1\e&NS,HMXB?\\
IGR J17464-3213 &17$^h$46$^m$24$^s$&32$^\circ$13\arcmin& 1979 & 6.0$\pm$0.1 (43$\pm$5)\d & 219 & 0.5\e     &BHC,LMXB\\
IGR J17597-2201 &17$^h$59$^m$42$^s$&22$^\circ$01\arcmin& 1588 & 6.9$\pm$0.2 & 179 & 2.8$\pm$0.4            &NS,B,D,LMXB\\
IGR/SAX J18027-2017 &18$^h$02$^m$42$^s$&20$^\circ$17\arcmin& 1274 & 4.4$\pm$0.2 & 146 & 5.3$\pm$0.5        &NS,P,HMXB?\\

\hline
\end{tabular}

\vspace{2mm}

\begin{list}{}
\item -- BHC -- black hole candidate, NS -- neutron star, HMXB - high mass
  X-ray binary, LMXB -- low mass X-ray binary, B -- burster, D-- dipper, P
  -- pulsar 
\item \a -- source coordinates as measured by INTEGRAL/IBIS
\item \b -- the effective exposure 
\item \c -- the average source flux in the energy band 18-60 keV, the 
flux was calculated assuming that the spectrum of the Crab nebula
has the shape $dN(E)=10.0 E^{-2.1} dE$
\item \d -- a maximum of the source flux in the 18-60 keV energy band is
given in parentheses
\item \e -- 3$\sigma$ upper limit
\end{list}
}
\end{table*}

The international gamma-ray observatory INTEGRAL (\cite{win03}) was launched
to the orbit with RUssian rocket PROTON from the cosmodrom Baikonur on Oct
17, 2002 (\cite{eis03}). It consists of four instruments designed for
investigation of galactic and extragalactic sources in a wide
X-ray/$\gamma$-ray energy band from 3 to 10000 keV: the gamma-telescope
IBIS, the SPI spectrometer, the X-ray monitor JEM-X and the optical monitor
OMC. In this paper we used data from the ISGRI detector (\cite{leb03}) of
the IBIS telescope (\cite{ube03}). Most sources were outside the JEM-X field
of view or were too faint for the detection with this instrument.

The SPI spectrometer has a relatively poor angular resolution
$\sim2-2.5$\deg\ (\cite{ved03}) that precludes its use for point sources
studies in such a crowded region as the Galactic Center. Therefore we do not
use data from the SPI spectrometer.

The Galactic Center region was observed with the INTEGRAL observatory from
August 23 till September 24, 2003 (Obs.IDs 0120213 and 0120016).  The total
exposure of these observations was slightly more than 2 Msec, but due to a
decrease of the IBIS efficiency for off-axis sources their effective
exposures were slightly lower (see Tab.\ref{tab1}).  In order to investigate
the variability of sources in hard X-rays on a time scale of half a year we
used a set of observations of the Galactic Center field performed by
INTEGRAL in March--April of 2004 with total exposure of $\sim250$ ksec
(Obs.ID 0220133).

The data of all IBIS/ISGRI observations were processed with the method
described by Revnivtsev et al. (2004a). In order to reconstruct the source
spectra from IBIS/ISGRI data we used a ratio of fluxes measured in different
energy channels to the fluxes measured by the ISGRI detector from the Crab
nebula in the same energy bands. Detailed analysis of Crab nebula
observations suggests that with the approach and software employed, the
conservative estimation of uncertainty in measurements of absolute fluxes
from the sources is about 10\% and the shape of the spectrum is about 5\%.
The last value was added to the following spectral analysis as a systematic
uncertainty in each energy channel.

The hard X-ray spectra were combined with spectra obtained with the PCA
spectrometer(3--20 keV) of the RXTE observatory (\cite{rxte}) where
possible. In particular, we used publically available data of the PCA
spectrometer of observations of IGR J17091-3624 (April 20, 2003, ObsID
80410-01-01-00 ) IGR J17464-3213 (from 26 Aug through 23 Sep, Obs. ID
80146-01), IGR/XTE J17391-3021 (28 Aug, Obs. ID 80074-01), IGR/XTE
J17597-2201 (June 3, Oct.  11, 2001, ObsID 60117-02). The PCA spectrometer
is sensitive to photons in the 3$-$20 keV energy band, its effective area is
$\sim6400$ cm$^2$ at energies 6$-$7 keV, the energy resolution is $\sim18$\%
at these energies.  For data reduction of the RXTE observations we used
standard programs of the FTOOLS 5.3 package. In order to trace the long term
variability of sources we have used the data of the all sky monitor (ASM) of
RXTE in the 1.3-12.2 keV energy band (http://xte.mit.edu/ASM\_lc.html).

\section*{Results}
\begin{table*}[t]
\centering

\caption{Best-fit parameters of source spectra in the \mbox{$>20$ keV}
  energy band\a \label{tab2}}    

\small{

\begin{tabular}{l|c|c|c}
\hline\hline
Source &\multicolumn{2}{c|}{Parameters} & $\chi^2$ (N d.o.f.) \\  \cline{2-3}
      & Photon index & Energy, keV &  \\
\hline
IGRJ 17091-3624 & 2.23$\pm$0.06 & 78\b  & 6.8 (10)\\
               & 1.59$\pm$0.07\c & 52\b   &10.1 (10)\\
IGR/XTE J17391-3021 & & 22$\pm$2\d  & 2.7 (6)\\
IGRJ 17464-3213 & 1.85$\pm$0.04 & 136\b    &11.2 (10)\\
IGRJ 17597-2201 & 1.70$\pm$0.07  & 78$\pm$3\e   &37.0 (32)\f\\
               & 3.05$\pm$0.47\c & &4.2 (5)\\
IGR/SAX J18027-2017 &           & 19$\pm$2\d & 1.1 (5)\\

\hline
\end{tabular}

\vspace{2mm}

\begin{list}{}
\centering
\item \a -- powerlaw, powerlaw with cutoff and bremsstrahlung models were
  used for the spectra approximations 
\item \b -- lower limits on the cutoff energy (2$\sigma$)
\item \c -- best-fit parameters of sources spectra obtained in 2004; for
  IGR/SAX J18027-2017 they are same as in 2003  
\item \d -- the temperature of the bremsstrahlung emission 
\item \e -- the energy of the exponential cutoff 
\item \f -- the combined RXTE and INTEGRAL fit  

\end{list}
}
\end{table*}

A list of sources with corresponding effective exposures and fluxes (in
mCrab units) measured in the 18-60 keV energy band is presented in Table
\ref{tab1} for 2003 and 2004. A flux of 1 mCrab in this energy band
corresponds to an energy flux of $\sim 1.4\times10^{-11}$ erg s$^{-1}$
cm$^{-2}$ for a source with a Crab-like spectrum with a photon index
$\Gamma=2.1$. Only three sources were detected in the last series of
observations; for the other two sources we obtained only upper limits. Most
of sources have a steady behavior, except for two objects, IGR J17464-3213
and IGR/XTE J17391-3021, which demonstrated flares during observations. In
Table \ref{tab1} we present both average and maximum values of their fluxes.
Model parameters of best-fit approximations of source spectra are listed in
Table \ref{tab2}.

Below we will discuss each source in detail.

\subsection*{IGR J17091-3624}

\begin{figure}[b]
\includegraphics[width=\columnwidth]{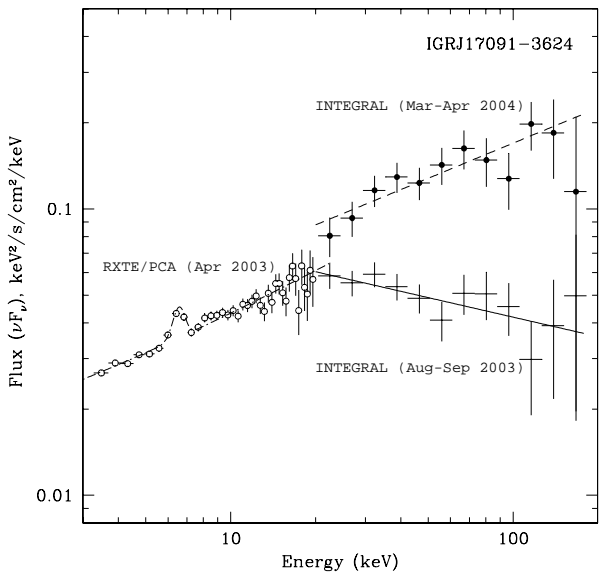}
\caption{The IGR J17091-3624 spectrum evolution in 2003-2004. Open circles
  represent RXTE data (Apr 2003), crosses -- INTEGRAL data (Aug-Sep 2003), 
  dark circles -- INTEGRAL data (Mar-Apr 2004). Corresponding best-fit
  approximations are shown by dash-dotted, solid and dashed lines,
  respectively.
  \label{fig:spc17091}}
\end{figure}

The source IGR J17091-3624 was discovered with the INTEGRAL observatory
during Galactic Center observations in April, 2003 (\cite{kuul03}). The
source demonstrated hard X-ray emission up to $\sim100$ keV with possible
softening of the spectrum during subsequent INTEGRAL observations.  Archival
data of MIR/KVANT/TTM and BeppoSAX/WFC showed that this source was detected
in soft X-rays from 1994 (\cite{rev03a}, \cite{intz03}). The region
containing a new object was almost immediately observed with the RXTE
observatory on 20 April 2003. Results of analysis of the RXTE data showed
that the source has a flux of $\sim 4$ mCrab in the 3-20 keV energy band;
its spectrum can be well approximated by a simple powerlaw model with a
photon index of $\Gamma=1.43$ (\cite{lut03a}). A power-density spectrum of
IGR J17091-3624 was described by the standard band-limited noise model
typical for X-ray binary systems in the hard/low spectral state. Based on
these results and taking into account the significant radio activity of IGR
J17091-3624 (\cite{rup03a}), Lutovinov \& Revnivtsev (2003) assumed that this
source is a possible black-hole candidate.

The source changed its brightness by a factor of $\sim2$ between April 2003,
Aug-Sept 2003 and April 2004 (see Table \ref{tab1} and \cite{kuul03}). Its
spectrum in the energy band 20-150 keV in 2003 can be described by a simple
powerlaw with a photon index $\Gamma=2.23$ (Table \ref{tab2},
Fig.\ref{fig:spc17091}), that is a softer than it was in RXTE observations.
As the source flux in the hard X-ray energy band increased in 2004, its
spectrum became harder; the photon index of the best fit power law
approximation is $\Gamma=1.59$ (Fig.\ref{fig:spc17091}). Lower limits on the
cutoff energy are $\sim78$ and $\sim52$ keV, respectively (Table
\ref{tab2}).  The value of $\Gamma$ obtained in the April 2004 observations
is close to those obtained with the RXTE observatory in 2003. The evolution
of the spectrum of IGR J17091-3624 during approximately one year of
observation is presented in Fig.\ref{fig:spc17091}. The spectrum of the
source obtained by the RXTE observatory in 2003 was taken from Lutovinov \&
Revnivtsev (2003).

\subsection*{IGR/XTE J17391-3021}
\begin{figure}[t]
\includegraphics[width=\columnwidth,bb=30 340 550 720]{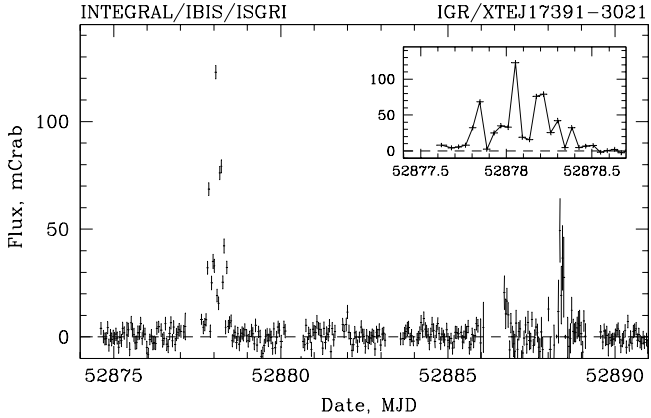}
\caption{INTEGRAL/IBIS/ISGRI light curve of IGR/XTE J17391-3021 in the 18-60
  keV energy band. Each point corresponds to a time bin of $\sim3500$ sec
  (one pointing), error bars correspond to 1$\sigma$. A detailed profile of
  the strong outburst on 26-27 Aug, 2003 is shown in the inset.
  \label{fig:lc17391}}
\end{figure}

On August 26, 2003 a strong outburst was detected by INTEGRAL during
observations of the Galactic Center field (\cite{sun03b}). Coordinates of a
possible new transient source IGR J17391-3021 were determined as ${\rm RA} =
17^h 39^m06^s$, ${\rm Dec} = -30$\deg$21$\arcmin$30$\arcsec with an accuracy
of $~$3\arcmin. The following analysis showed that in spite of
$\sim5.5$\arcmin\ separation, the newly discovered source is likely to be
the same transient source (XTE J1739-302) that was seen by the RXTE in 1998
(\cite{smi98}). Therefore in the subsequent analysis and discussions we
called this source the double name IGR/XTE J17391-3021.

Observations of this field performed with the Chandra observatory and VLA
revealed inside the INTEGRAL coordinate error box two sources with
non-overlapping coordinates: ${\rm RA} = 17^h 39^m11^s.58$, ${\rm Dec} =
-30$\deg$20$\arcmin$37.6$\arcsec (Chandra, an accuracy is better than a few
arcsec, \cite{smi03}) and ${\rm RA} = 17^h 39^m01^s.52$, ${\rm Dec} =
-30$\deg$19$\arcmin$34.9$\arcsec (VLA, an accuracy of 0.2\arcsec,
\cite{rup03b}). The distance between these two positions is quite large,
indicating that the radio source may be not related to IGR/XTE J17391-3021.

The light curve of the source IGR/XTE J17391-3021 obtained with the INTEGRAL
observatory in August-September, 2003 is presented in Fig.\ref{fig:lc17391}.
The total length of the outburst was less than one day; its structure was
complex and included several peaks. The maximal flux detected in the 18-60
keV energy band is $\sim$120 mCrab. The follow-up observations performed
with the RXTE observatory on Aug 28 gave us only an upper limit to the
source flux of $\sim3$ mCrab in the 3-20 keV energy band. Thus the radio
flux observed on this day by VLA in the INTEGRAL error circle
(\cite{rup03b}) may be the reflection of the previous X-ray activity of the
source or may be not related to IGR/XTE J17391-3021.

\begin{figure}[t]
\includegraphics[width=\columnwidth]{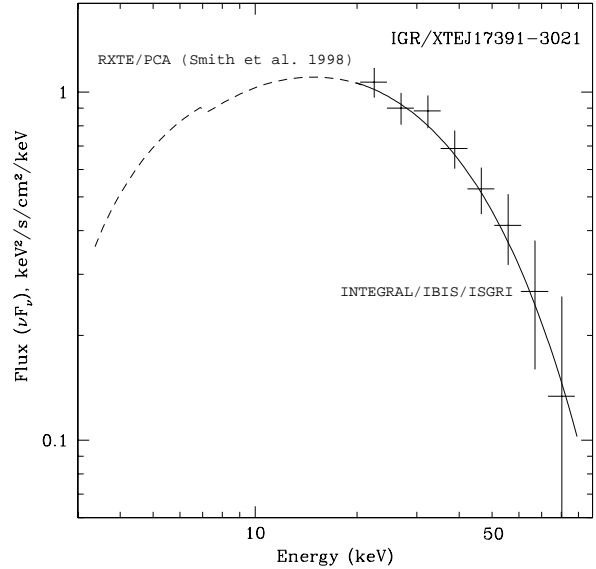}
\caption{The spectrum of IGR/XTE J17391-3021, measured with INTEGRAL
  (crosses) and its best-fit approximation (solid line). The best-fit
  approximation model of the source spectrum obtained with RXTE in 1998 is
  shown by a dashed line. We fixed its parameters to values obtained by
  Smith et al.(1998) and fitted only its normalization to the flux measured
  with INTEGRAL.  \label{fig:spc17391}}
\end{figure}

We reconstructed the spectrum of the source during its different flux values
and found that the shape of the spectrum was practically identical. In the
following analysis we used the energy spectrum of the source in the maximum
of the outburst (Fig.\ref{fig:spc17391}).  Approximation of this spectrum by
a single power law (with $\Gamma\sim3$) gives an unacceptably high $\chi^2$
value (10.8 for 5 d.o.f.). An optically thin thermal bremsstrahlung model or
power law model with exponential cutoff provides much better fits (see Table
\ref{tab2}). Such a spectral shape in hard X-rays is typical for of binary
systems with neutron stars. The fit temperature was found to be the same as
measured by Smith et al. (1998). These authors also reported that timing
analyses of the source performed with the RXTE observatory have not show
pulsations with an amplitude above 2\% with a period less than 300 sec.
Also no X-ray burst-like events were observed from the source.  Thus the
nature of IGR/XTE J17391-3021 is still unclear; however the observed
spectral shape indicates that it is a neutron star binary, and very likely a
high mass X-ray binary.

During INTEGRAL observations one more outburst with a complex structure was
detected from IGR/XTE J17391-3021 (see Fig.\ref{fig:lc17391}). The maximum
flux was $\sim$2 times lower than that observed on 26-27 Aug; the shape of
the spectrum was the same.

\subsection*{IGR J17464-3213 (= XTE J17464-3213 = H 1743-322)}

The transient source IGR J17464-3213 was detected by the INTEGRAL
observatory on March 21, 2003 (\cite{rev03b}). During the next several days
the source flux increased by a factor of $\sim3$ and reached the value of
$\sim60-70$ mCrab in a wide energy band of 15-200 keV. Subsequent
investigations (\cite{mar03a}) showed that the position of the source is
consistent with one of two equally possible positions of the source
H1743-322, which was detected in 1977-1978 with \mbox{HEAO-1}
(\cite{gur78}). The follow-up observations with VLA revealed a probable
radio counterpart of IGR J17464-3213 from which a strong radio flare was
detected on April 8, 2003 (\cite{rup03c}). The Magellan optical observations
gave only upper limits to the possible conterpart in R and J filters
(\cite{ste03}). It was proposed that the source IGR J17464-3213 observed by
INTEGRAL is a recurrent appearance of the source H1743-322 (\cite{par03}).
The outburst lasted more than 200 days. Special TOO observations were
performed with the INTEGRAL observatory during the rising part of the
outburst and near its maximum (\cite{par03}). These observations showed that
the source demonstrated a behavior typical of X-ray novae.  Its spectrum in
a wide energy band of 4-200 keV was well approximated by a simple powerlaw
model with a photon index of $\sim2.7$ with indications of a soft component,
that is typical for the high/soft state of X-ray novae. The source
monitoring performed with RXTE in May-July, 2004 allowed to observe it in
several black hole states and revealed various types of variability,
including QPOs (\cite{hom04}).

\begin{figure}[t]
\includegraphics[width=\columnwidth]{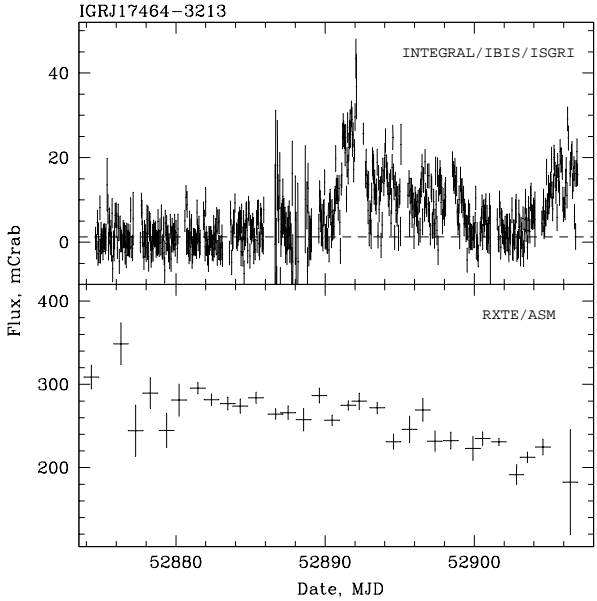}
\caption{Light curves of IGR J17464-3213 obtained in August-September 2003
  with INTEGRAL/IBIS in the 18-60 keV energy band (upper panel) and RXTE/ASM
  in the 1.3-12.2 keV energy band (bottom panel). Each point in INTEGRAL
  data corresponds to a time bin of $\sim3500$ sec, in RXTE data -- to the
  one-day averaging. Error bars corresponds to 1$\sigma$.
  \label{fig:lc17464}}
\end{figure}

\begin{figure}[t]
\includegraphics[width=\columnwidth]{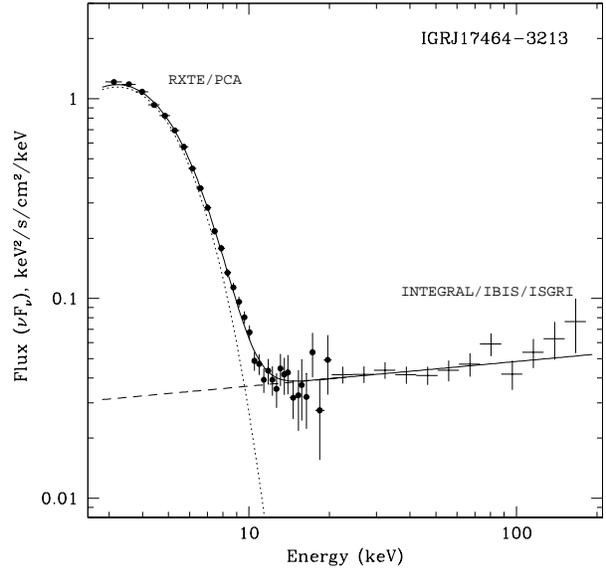}
\caption{Broadband spectrum of IGR J17464-3213 obtained with INTEGRAL
  (crosses) and RXTE (dark circles) observatories. Best-fit approximation
  with the blackbody+powerlaw model is presented by solid line,
  contributions of each component are shown by dotted and dashed lines,
  respectively.  \label{fig:spc17464}}
\end{figure}

At the moment of our observations in August-September 2003 the source flux
in the RXTE/ASM energy band (1.3-12.2 keV) dropped more than 2 times its
maximal value and was still decreasing (see also \cite{hom04}). In the hard
energy band (18-60 keV) the source flux was practically constant at the
level of $\sim1.5$ mCrab during the first part of the observations and
demonstrated outburst-like events with a maximum of $\sim40$ mCrab during
the second one (Fig.\ref{fig:lc17464}, \cite{greb03}). The spectra of the
source in the hard X-ray energy band did not change during these flux
variations. In order to construct a broadband X-ray spectrum of the source
we combined data of INTEGRAL and RXTE observatories
(Fig.\ref{fig:spc17464}). The spectral points obtained with the RXTE/PCA
were rescaled to match those of INTEGRAL/IBIS. The best-fit model of the
broadband spectrum consists of two components: 1) a soft component which can
be described by the blackbody disk emission (\cite{ss73}) with the
temperature $kT_{in}=1.05\pm0.01$ keV; 2) a hard powerlaw tail with
$\Gamma=1.85\pm0.04$. This tail was visible up to $\sim200$ keV without any
cutoff (2$\sigma$ lower limit on the cutoff energy is $\sim136$ keV, see
Table \ref{tab2}).

Note that for our 2004 observations (April 2004) the outburst from IGR
J17464-3213 was finished and we obtained only an upper limit on its flux
(Table \ref{tab1}).

Summarizing all the above we can conclude that IGR J17464-3213 is a
classical X-ray novae -- a low mass X-ray binary harboring a black hole --
which experienced a recurrent outburst in 2003. 

\subsection*{IGR J17597-2201}

The source IGR J17597-2201 was detected by INTEGRAL on March 30 -- April 1,
2003 with a flux of $\sim5$ mCrab in the 15-40 keV energy band
(\cite{lut03b}). Two weeks later the source intensity increased to 10-15
mCrab in the same energy band. Just after the discovery of this new
transient source, Markwardt \& Swank (2003b) reported that its position is
consistent with a transient XTE J1759-220 which had been observed with RXTE
since Feb, 2001, but this information was not published. The regular
galactic bulge scans performed with the RXTE showed that the source
demonstrates a strong erratic flux variability in the standard X-ray band.
There was an observational evidence of dipping behavior that implies high
inclination of the binary. Also there were observed X-ray bursts, suggesting
that the source is a neutron star binary. No QPO or pulsations were detected
(\cite{mar03b}).
\begin{figure}[t]
\includegraphics[width=\columnwidth]{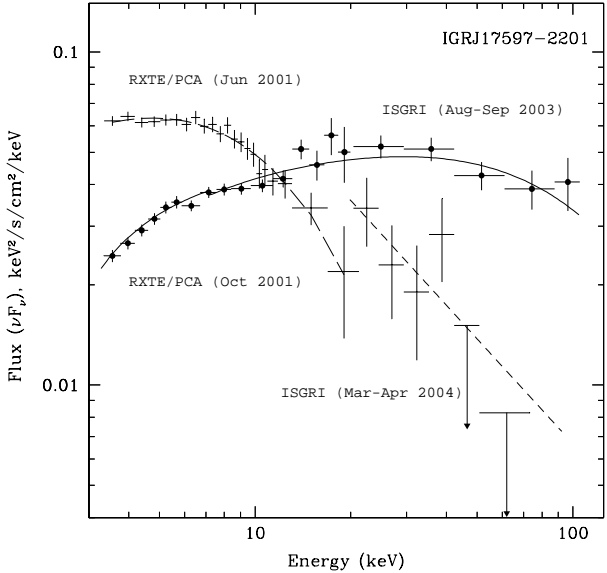}
\caption{IGR J17597-2201 spectrum changes observed by RXTE and INTEGRAL 
  observatories in 2001 and 2003-2004, respectively. Crosses represent
  the high/soft state of the source, dark circles -- low/hard state.
  Corresponding best-fit approximations are shown by dashed and solid
  lines, respectively. The contribution of the galactic ridge to
  RXTE/PCA spectra was subtracted.\label{fig:spc17597}}
\end{figure}

IGR J17597-2201 was observed continuously during both series of our Galactic
Center field observations. The source flux was $\sim 2.5$ times higher in
2003 than in 2004 (see Table \ref{tab1}). Such a change in the source flux
was accompanied by a strong change of its spectral shape
(Fig.\ref{fig:spc17597}). During 2003 observations the source was hard
($\Gamma\sim2.2$) and its flux was detected up to $\sim100$ keV; during 2004
the source spectrum was much softer (see Fig.\ref{fig:spc17597}). Such a
difference can be interpreted as a spectral transition from the hard/low to
the soft/high state of the neutron star binary. As such a transition
typically occurs in the luminosity range around $\sim 0.01 L_{\rm Edd}\sim
10^{36}$ erg s$^{-1}$ (see e.g. \cite{mac03} and references therein), we can
roughly (within a factor of $\sim2$) estimate the distance to the source
$d\sim5-10$ kpc, that may confirm that IGR J17597-2201 is located in the
Galactic bulge.

In order to obtain a broadband energy spectrum of the source we used data
from the RXTE archive (observations performed on June 3, 2001 and Oct. 11,
2001). For the reconstruction of the IGR J17597-2201 spectrum we should
carefully take into account the contribution of the Galactic ridge emission
to the flux detected by the RXTE/PCA (similar to what was done in Revnivtsev
2003a). To do this we predicted the spectrum of the Galactic ridge at the
position of IGR J17597-2201 using the following assumptions. We have taken
the spectral shape of the Galactic ridge emission in the form described by
Revnivtsev (2003b). We assumed that most of the line blend emission
($\sim$6-7 keV) detected by the RXTE/PCA at the position of IGR J17597-2201
comes from the Galactic ridge component. This assumption agrees well with
parameters of the ridge emission presented in Valinia \& Marshall (1999) or
Revnivtsev (2003b) and allow us to relatively accurately predict the flux of
the Galactic ridge at the position of the source. The predicted Galactic
ridge component was subtracted from the spectra observed by RXTE/PCA.
Resulting broadband spectra are presented in Fig.\ref{fig:spc17597}.

The combined (INTEGRAL and RXTE) spectrum of IGR J17597-2201 in the hard
state can be well described by a power law with an exponential cutoff at
high energies -- a model typical of neutron star binaries in the low/hard
spectral state (see Table \ref{tab2}). The soft/high state spectrum of the
source can be well approximated (due to limited statistics) by a thermal
bremsstrahlung model with the temperature $kT=7.5\pm0.5$ keV. Such a
spectral shape resembles those of neutron star binaries in their soft/high
spectral state and the temperature parameter $kT$ strongly exceeds those of
black holes binaries in their soft/high spectral state (sf. IGR J17464-3213
on Fig.\ref{fig:spc17464}). This conclusion support the suggestion of
Markwardt \& Swank (2003b) about the nature of IGR J17597-2201.

\subsection*{IGR/SAX J18027-2017}
\begin{figure}[t]
\includegraphics[width=\columnwidth]{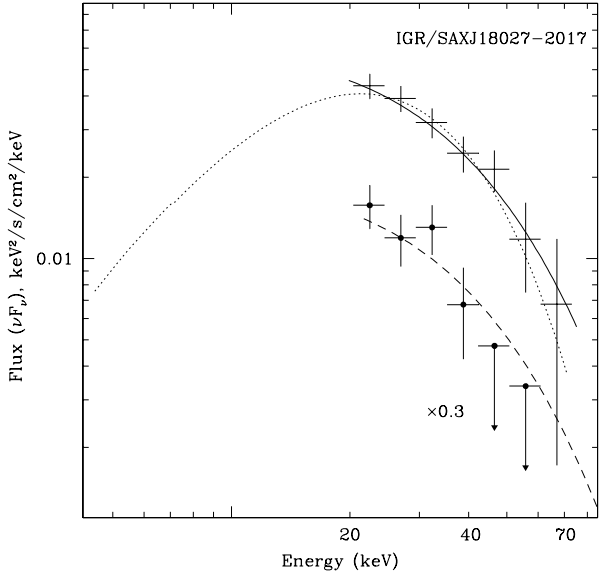}
\caption{Spectra of IGR/SAX J18027-2017 measured with INTEGRAL in 2003
  (crosses) and 2004 (filled circles, scaled by a factor of 0.3). Solid and
  dashed lines represent corresponding best-fit approximations. An
  extrapolation of the spectral model obtained by BeppoSAX to a wide energy
  band is shown by dotted line (see text for detail).\label{fig:spc18027}}
\end{figure}

First information about a detection of a possible new source IGR J18027-2017
near the bright X-ray source GX9+1 with the INTEGRAL observatory was
discussed in April 2003 during INTEGRAL Core Program observations of the
Galactic Center field. An analysis of the archival data of the BeppoSAX
observatory revealed a source in the Galactic Center, which was named SAX
J1802.7-2017 (\cite{aug03}). The detailed timing analysis produced by
Augello et al. (2003) showed that this source is an X-ray pulsar with a
pulse period of $\sim139.6$ sec and allowed one to determine binary
parameters of the system.

IGR/SAX J18027-2017 was within the IBIS field of view during all the
Galactic Center observations in 2003 and 2004. The flux registered from the
source in the 18-60 keV energy band was nearly constant and equal in both
sets of observations (see Tab.\ref{tab1}). The shape of the source spectrum
in hard X-rays did not change during observations. It can be described by a
power law model with an exponential cutoff at high energies.  The data of
the IBIS telescope very poorly constrain the photon index of the powerlaw,
but we can determine the energy of the cutoff $E_{\rm cut} \sim18$ keV, that
is typical for X-ray pulsars. In order to present a broadband spectrum of
the source (Fig. \ref{fig:spc18027}), we have fitted the spectrum observed
by IBIS/ISGRI with a typical pulsar spectral model (power law with
exponential cutoff at high energies) while fixing the power law index on the
value obtained by Augello et al. (2003) in the soft energy band.  The
best-fit value of the cutoff energy of $\sim10$ keV resulting from such a
procedure is lower than that obtained by fitting only hard X-ray data.

\section*{Summary}

We presented an analysis of INTEGRAL observations of five newly discovered
sources -- IGR J17091-3624, IGR/XTE J17391-3021, IGR J17464-3212, IGR
J17597-2201 and SAX/IGR J18027-2017.  Taking into account the mentioned
above information about the sources we can conclude that all of them have a
Galactic origin.

The nature of these sources can be summarized as following:

\begin{itemize}
\item  IGR J17091-3624 is a likely black hole candidate
  
\item properties of IGR/XTE J17391-3021 indicate that it is likely a
  high-mass neutron star binary. It is possible that it can be an accreting
  X-ray pulsar

\item IGR J17464-3212 is a classical X-ray novae harboring a black hole,
  likely a low mass X-ray binary

\item IGR J17597-2201 is a low mass X-ray binary, harboring a neutron star.

\item SAX/IGR J18027-2017 is a neutron star binary, accreting X-ray pulsar,
likely HMXB

\end{itemize}

\begin{acknowledgements}
  The authors thank Eugene Churazov for developing of methods of analysis of
  the IBIS data and software. We would like to thank Marat Gilfanov and
  referee, Arvind Parmar, for useful comments. This research has made use of
  data obtained through the INTEGRAL Science Data Center (ISDC), Versoix,
  Switzerland, Russian INTEGRAL Science Data Center (RSDC), Moscow, Russia,
  and High Energy Astrophysics Science Archive Research Center Online
  Service, provided by the NASA/Goddard Space Flight Center. This work was
  partially supported by grants of Minpromnauka NSH-2083.2003.2 and
  40.022.1.1.1102 and the program of Russian Academy of Sciences
  ``Non-stationary phenomena in astronomy''. AL, MR and SM acknowledge the
  support of RFFI grant 04-02-17276 and the International Space Science
  Institute (ISSI, Bern).

\end{acknowledgements}

\end{document}